# PROBING LEPTON-NUMBER-VIOLATING COUPLINGS OF DOUBLY CHARGED HIGGS BOSONS AT AN $e^-e^-$ COLLIDER[1]

J.F. Gunion

*Davis Institute for High Energy Physics*
*University of California, Davis, CA 95616*

## Abstract

The doubly charged Higgs bosons $\Delta^{--}$ that are present in exotic Higgs representations can have lepton-number-violating couplings to $e^-e^-$. We discuss general constraints and phenomenology for the $\Delta^{--}$ and demonstrate that extremely small values for the $e^-e^- \to \Delta^{--}$ coupling (some 8 orders of magnitude smaller than the current limit) would produce observable signals for $\Delta^{--}$ production in direct $s$-channel production at an $e^-e^-$ collider.

## 1 Introduction

Doubly-charged scalar particles abound in exotic Higgs representations and appear in many models [1, 2, 3]. For example, a Higgs doublet representation with $Y = -3$ contains a doubly-charged $\Delta^{--}$ and a singly-charged $\Delta^-$. If part of a multiplet with a neutral member, a $\Delta^{--}$ would immediately signal the presence of a Higgs representation with total isospin $T = 1$ or higher. Most popular are the complex $Y = -2$ triplet Higgs representations, such as those required in left-right symmetric models, that contain a $\Delta^{--}$, a $\Delta^-$ and a $\Delta^0$.

Of course, in assessing the attractiveness of a Higgs sector model containing a $\Delta^{--}$ many constraints need to be considered. For triplet and higher representations containing a neutral member, limits on the latter's vacuum expectation value required to maintain $\rho = 1$ are generally severe. (The first single representation beyond $T = 1/2$ for which $\rho = 1$ is automatic regardless of the vev is $T = 3, Y = -4$, whose $T_3 = 0$ member is doubly-charged.) Models with $T = 1$ and $T = 2$ can 'automatically' have $\rho = 1$ at tree-level by combining representations (the most well-known example being a Higgs sector containing both $Y = 0$ and $Y = -2$ triplets whose neutral members have the same vacuum expectation value). However, such models generally require fine-tuning in order to preserve $\rho = 1$ at one-loop. The simplest way to avoid such problems is to either consider representations that simply do not have a neutral member (for example, a $Y = -3$ doublet or a $Y = -4$ triplet representation), or else models in which the vacuum expectation value of the neutral member is precisely zero. We will only consider models of this type in what follows.

---

[1]To appear in the *Proceedings of the Santa Cruz Workshop on $e^-e^-$ Physics at the NLC*, University of California, Santa Cruz, Sept. 4-5, 1995.

Another source of motivation for and constraints on Higgs representations arises if we require unification of the coupling constants without intermediate scale physics. In the Standard Model, it is worth noting that quite precise unification is possible for a relatively simple Higgs sector that includes a triplet Higgs representation — namely, a single $|Y|=2$ triplet in combination with either one or two $|Y|=1$ doublets (the preferred number of doublets depends upon the precise value of $\alpha_s(m_Z)$). The neutral member of this single triplet would need to have zero vacuum expectation value to avoid $\rho$ problems. In the case of the minimal supersymmetric extension of the Standard Model, the only Higgs sector that gives precise unification is that comprised of exactly two doublet Higgs representations (plus possible singlet representations); any extra doublet representations (including ones with a doubly-charged boson) or any number of triplet or higher representations (whether or not *any* doublets are present) would destroy unification. Thus, doubly-charged Higgs bosons would seem to be very unlikely in this context. However, by including appropriate intermediate scale physics, supersymmetric models with triplet and higher representations can often be made consistent with unification. In particular, supersymmetric extensions of the left-right symmetric models, which must contain triplet Higgs representations, typically do have matter at intermediate scales and the exact masses can comfortably be adjusted to achieve coupling unification.

Thus, allowing for the possibility that the two-doublet MSSM might not be nature's choice, experimentalists should be on the look-out for signatures of exotic Higgs representations. Since many of the more attractive higher representations include a doubly-charged Higgs boson, it is important to consider how to search for and study such a particle. The phenomenology of the $\Delta^{--}$ derives from its couplings. Tri-linear couplings of the type $W^-W^- \to \Delta^{--}$ are not present in the absence of an enabling non-zero vacuum expectation value for the neutral member (if present) of the representation, and $q'\bar{q}\Delta^{--}$ couplings are obviously absent. There are always couplings of the form $Z, \gamma \to \Delta^{--}\Delta^{++}$, and these can be useful for production of the $\Delta^{--}$, as outlined later. However, an especially interesting possibility is the lepton-number-violating $e^-e^- \to \Delta^{--}$ coupling that is sometimes allowed by symmetry. For $Q = T_3 + \frac{Y}{2} = -2$ the allowed cases are:

$$\begin{aligned} e_R^- e_R^- &\to \Delta^{--}(T_3=0, Y=-4)\,,\\ e_L^- e_R^- &\to \Delta^{--}(T_3=-\tfrac{1}{2}, Y=-3)\,,\\ e_L^- e_L^- &\to \Delta^{--}(T_3=-1, Y=-2)\,. \end{aligned} \qquad (1.1)$$

Note that the above cases include the $T=3, Y=-4$ representation that yields $\rho=1$, the $T=1/2, Y=-3$ doublet and $T=1, Y=-4$ triplet representations with no neutral member, and the popular $T=1, Y=-2$ triplet representation.

The above list of allowed couplings is expanded in the left-right symmetric models where $Q = T_3^L + T_3^R + \frac{Y}{2}$. Indeed, in left-right symmetric models there is a $|Y| = 2$ Higgs triplet representation that has lepton-number-violating couplings to right-handed leptons. The right-handed neutrino states acquire a large mass when the neutral member of this 'right-handed' triplet acquires a non-zero vev. This large mass in turn leads to a light neutrino mass eigenstate via the popular see-saw mechanism. The phenomenological analysis we present below would have to be extended in the case of this triplet since its neutral member has a vev. However, in the left-right symmetric models there is a second, 'left-handed', $|Y| = 2$ triplet that couples to left-handed neutrinos and leptons. The strength of the coupling is the same as that associated with the right-handed sector. As usual, the neutral member of this 'left-handed' triplet must have a very small vev in order to preserve $\rho = 1$, and the phenomenology of this triplet's doubly-charged member would be as described below. The strength of the lepton-number-violating coupling (common to the right and left sectors) required to make the see-saw mechanism work properly in the right-handed sector is typically such as to fall into the range that we shall claim can be probed in $e^-e^-$ collisions through production of the doubly-charged member of the left-handed triplet.

In the case of a $|Y| = 2$ triplet representation (to which we now specialize) the lepton-number-violating coupling to (left-handed) leptons is specified by the Lagrangian form:

$$\mathcal{L}_Y = i h_{ij} \psi_{iL}^T C \tau_2 \Delta \psi_{jL} + \text{h.c.}, \tag{1.2}$$

where $i, j = e, \mu, \tau$ are generation indices, the $\psi$'s are the two-component left-handed lepton fields ($\psi_{\ell L} = (\nu_\ell, \ell^-)_L$), and $\Delta$ is the $2 \times 2$ matrix of Higgs fields:

$$\Delta = \begin{pmatrix} \Delta^+/\sqrt{2} & \Delta^{++} \\ \Delta^0 & -\Delta^+/\sqrt{2} \end{pmatrix}. \tag{1.3}$$

In left-right symmetric models $\Delta$ would have to be subscripted as $\Delta_L$, and there would be a Lagrangian component analogous to that given in Eq. 1.2 with $L \to R$ everywhere.

The strengths of the couplings in Eq. 1.2 are specified by the $h_{ij}$; $e^-e^- \to \Delta^{--}$ will be controlled by $h_{ee}$. Limits on the $h_{ij}$ come from many sources. Experiments that directly place limits on the $h_{ij}$ by virtue of the $\Delta^{--} \to \ell^-\ell^-$ couplings include Bhabbha scattering, $(g-2)_\mu$, muonium-antimuonium conversion, and $\mu^- \to e^-e^-e^+$. For some details and references see, for example, Refs. [2, 4]. One finds limits of

$$\begin{aligned} |h_{ee}^{\Delta^{--}}|^2 &\lesssim 10^{-5} m_{\Delta^{--}}^2(\text{GeV}) \\ |h_{\mu\mu}^{\Delta^{--}}|^2 &\lesssim 4 \times 10^{-5} m_{\Delta^{--}}^2(\text{GeV}) \\ |h_{ee}^{\Delta^{--}} h_{\mu\mu}^{\Delta^{--}}| &\lesssim 6 \times 10^{-5} m_{\Delta^{--}}^2(\text{GeV}) \\ |h_{e\mu}^{\Delta^{--}} h_{ee}^{\Delta^{--}}| &\lesssim 5 \times 10^{-11} m_{\Delta^{--}}^2(\text{GeV}) \end{aligned} \tag{1.4}$$

from the above respective sources. (In our notation, $h_{ij}^{\Delta^{--}}$ refers explicitly to the $h_{ij}$ couplings as they determine $\Delta^{--}$ interactions.) The last of these limits clearly suggests small off-diagonal couplings, and in what follows we shall assume that the couplings are entirely diagonal. We shall adopt the conventional form for these couplings of

$$|h_{\ell\ell}^{\Delta^{--}}|^2 \equiv c_{\ell\ell} m_{\Delta^{--}}^2 \, (\text{GeV}), \tag{1.5}$$

where it will be useful to keep in mind that $c_{ee} \lesssim 10^{-5}$ is the strongest of the limits.

Finally, we remark that constraints on the $h_{ij}$ through couplings for the $\Delta^0$ and $\Delta^-$ should also generally be incorporated for the specific model being considered. The only such constraint that is potentially stronger than those outlined above is that associated with Majorana mass terms for the neutrinos coming from their couplings to the $\Delta^0$. One finds [3] $m_{ij} = 2h_{ij}\langle\Delta^0\rangle$. If $\langle\Delta^0\rangle$ were of order the SM vev, then, for example, $m_{\nu_e} \lesssim 1\,\text{eV}$ (as required to prevent neutrinoless double-beta decay from being observable) would imply $h_{ee} \lesssim 10^{-14}$. However, this constraint clearly goes away if $\Delta^0$ does not have a significant vev. In the present study, we assume (as stated earlier) that the vev is, in fact, exactly zero, so as to avoid problems associated with maintaining $\rho = 1$ naturally.

## 2 General Phenomenology for a $\Delta^{--}$

Given adequate machine energy, production of $\Delta^{--}\Delta^{++}$ via $\gamma, Z$ exchange at either an $e^+e^-$ or $pp$ collider will yield an observable signal. At an $e^+e^-$ collider the cross section for pair production of a boson with weak isospin $T_3$ and charge $Q$ and its conjugate is given at energies $s \gg m_Z^2$ by:

$$\sigma^{\text{pair}} = \left(\frac{4}{3}\frac{\pi\alpha^2}{s}\right)\frac{2\beta^3}{\sin^4 2\theta_W}\left\{\left(\frac{1}{2}T_3 + \sin^2\theta_W\left[\frac{1}{2}Q - T_3\right]\right)^2 + (Q - T_3)^2\sin^4\theta_W\right\}. \tag{2.1}$$

For the case of a $T_3 = -1, Y = -2$ $\Delta^{--}$ we find $\{\ldots\} \to \frac{1}{4} + \sin^4\theta_W$ and $\sigma^{\text{pair}} \sim 488\,\text{fb}\beta^3$ at $\sqrt{s} = 500\,\text{GeV}$, which yields about 11 fb for $m_{\Delta^{--}} \sim 240\,\text{GeV}$. In other words, we would have at least 50 events for $L = 50\,\text{fb}^{-1}$ for masses up to within 10 GeV of threshold. We shall discuss signatures in a moment, but this number of events would generally be adequate for $\Delta^{--}\Delta^{++}$ detection. However, the $e^+e^- \to \Delta^{--}\Delta^{++}$ process has a crucial limitation. It allows detection of the $\Delta^{--}$ only up to $m_{\Delta^{--}} \lesssim \sqrt{s}/2$. This is only half the kinematical reach of $s$-channel production of the $\Delta^{--}$ in the $e^-e^-$ mode of operation at the same $\sqrt{s}$ value. Further, detection of the $\Delta^{--}$ prior to the construction and operation of the $e^+e^-, e^-e^-$ collider NLC complex would be very important in determining the energy range over which good luminosity and good energy resolution for $e^-e^-$ collisions should be a priority. Thus, it is fortunate that observation of $\Delta^{--}\Delta^{++}$ pairs with $m_{\Delta^{--}} \sim 500\,\text{GeV}$ is straightforward at the LHC. The mass reach for pair production at a $pp$ collider increases rapidly with machine energy. At the LHC, $\sigma^{\text{pair}} \sim 1\,\text{fb}$ at $m_{\Delta^{--}} = 800\,\text{GeV}$, the precise number depending upon $T_3$. With $L = 100\,\text{fb}^{-1}$, there would clearly be a large number of $\Delta^{--}\Delta^{++}$ events for any $\Delta^{--}$ with $m_{\Delta^{--}} \lesssim 500\,\text{GeV}$. For sufficiently small $m_{\Delta^{--}}$, observation of $\Delta^{--}\Delta^{++}$ pair production might even be possible at the Tevatron.

Another potential production mechanism for the $\Delta^{--}$ at $pp$ colliders is the fusion process, $W^-W^- \to \Delta^{--}$. However, the required tri-linear coupling is zero given our assumption that the vev of the neutral member (if there is one) of the Higgs representation is zero (thereby avoiding naturalness problems associated with maintaining $\rho = 1$). A general discussion of event rates for $W^-W^- \to \Delta^{--}$ fusion for typical models in which the vev is not zero can be found in Refs. [2, 3, 1].

Decays of a $\Delta^{--}$ are generally quite exotic [2, 3]. If there is an enabling non-zero vev, then $\Delta^{--} \to W^-W^-$ decays can be very important. If this coupling is absent (as we assume), then possible two-body decays include $\Delta^{--} \to \Delta^-W^-$, $\Delta^{--} \to \Delta^-\Delta^-$ and, if the lepton coupling is present, $\Delta^{--} \to \ell^-\ell^-$. Assuming some reasonable amount of degeneracy of the masses of different members of the multiplet, the $\Delta^{--} \to \Delta^-\Delta^-$ decay is likely to be disallowed. Thus, we will focus on the $\Delta^-W^-$ and $\ell^-\ell^-$ final states. (Generalization of our discussion if other decays are present will be apparent.) For a $T=1, Y=-2$ triplet we find (see, for example, Refs. [2, 3])

$$\Gamma(\Delta^{--} \to \Delta^-W^-) = \frac{g^2}{16\pi}\frac{m^3_{\Delta^{--}}\beta^3_{\Delta^-W^-}}{m^2_W}\,, \quad \Gamma(\Delta^{--} \to \ell^-\ell^-) = \frac{\left|h^{\Delta^{--}}_{\ell\ell}\right|^2}{8\pi}m_{\Delta^{--}}\,, \tag{2.2}$$

where $\beta_{\Delta^-W^-}$ is the usual phase space suppression factor. It is convenient to rewrite these widths, using Eq. 1.5, in the forms:

$$\begin{aligned}\Gamma(\Delta^{--} \to \Delta^-W^-) &= (1.3\,\text{GeV})\left(\tfrac{m_{\Delta^{--}}}{100\,\text{GeV}}\right)^3 \beta^3_{\Delta^-W^-}\,,\\ \Gamma(\Delta^{--} \to \ell\ell) &= (0.4\,\text{GeV})\left(\tfrac{c_{\ell\ell}}{10^{-5}}\right)\left(\tfrac{m_{\Delta^{--}}}{100\,\text{GeV}}\right)^3\,.\end{aligned} \tag{2.3}$$

In order to gain a rough idea of the relative magnitude of these widths, consider the case [2] $m_{\Delta^{--}} = 360\,\text{GeV}$, $m_{\Delta^-} = 250\,\text{GeV}$. From Eq. 2.3 we find $\Gamma(\Delta^{--} \to \Delta^-W^-) \sim 2\,\text{GeV}$ and $\Gamma(\Delta^{--} \to \ell^-\ell^-) = 19\,\text{GeV}\left(\tfrac{c_{\ell\ell}}{10^{-5}}\right)$. This makes it clear that if any $c_{\ell\ell}$ is near $10^{-5}$ then that $\ell^-\ell^-$ mode is very likely to have a partial width larger than the $\Delta^-W^-$ partial width. Since there are currently no limits on $c_{\tau\tau}$, the $\tau^-\tau^-$ channel could easily have the largest partial width and be the dominant decay of the $\Delta^{--}$. On the other hand, when we discuss probing very small $c_{ee}$ values, we must keep in mind that if the other $c$'s are of similar size then the $\Delta^-W^-$ mode is quite likely to be dominant if it is kinematically allowed.

The implications for detection of $\Delta^{--}\Delta^{++}$ pairs in $e^+e^-$ or $pp$ collisions are obvious. If one or more of the $c_{\ell\ell}$'s is $\gtrsim 10^{-5}$, the $\ell^-\ell^-$ channel with the largest $c_{\ell\ell}$ will dominate $\Delta^{--}$ decays. For $\ell = e$ or $\mu$, we will have spectacular signatures of two like-sign lepton pairs of equal mass. Even a very few events of this type will constitute an unambiguous signal. If it is $c_{\tau\tau}$ that is largest, the $4\tau$ final state would have four energetic leptons and/or isolated pions plus missing energy a large fraction of the time and be clearly distinct from possible backgrounds. If *all* the $c_{\ell\ell}$'s are small and the $\Delta^-W^-$ mode is allowed, then since the $\Delta^-$ would most probably decay via $\Delta^- \to ZW^-, \Delta^0 W^-$, we would have final states containing two $W^-$'s, two $W^+$'s, and $ZZ$, $Z\Delta^0$, or $\Delta^0\Delta^0$. Only a fraction $(2/9)^4 = 0.0025$ of the time would all the $W$'s decay to $\ell = e$ or $\mu$; although this is a very background-free channel, the event rate would not generally be adequate. Reconstruction in hadronic channels of some of the $W$'s would be necessary. Still, at least one or two leptons could be required, along with pairs of energetic jets having mass $m_W$, and a viable signal is likely to emerge from a sample of 100 or more $\Delta^{--}\Delta^{++}$ pair events at the LHC. Thus, as stated earlier, we believe it is entirely reasonable to suppose that a $\Delta^{--}$ in the $m_{\Delta^{--}} \lesssim 500\,\text{GeV}$ mass range relevant for a $\sqrt{s} = 500\,\text{GeV}$ $e^-e^-$ collider would already have been observed at the LHC, regardless of the magnitude of the $c_{\ell\ell}$'s.

If a $\Delta^{--}$ is found, we shall certainly want to learn all about it. However, only limited information concerning the $c_{\ell\ell}$'s will be available from the $\Delta^{--}\Delta^{++}$ pair production process. The most optimistic scenario is that in which the $\Delta^{--} \to \Delta^-W^-$ decay channel is observed, and yet some $4\ell$ final state has a significant branching ratio in $\Delta^{--}\Delta^{++}$ pair production. (This would imply that the corresponding $c_{\ell\ell}$ is at or above the $10^{-5}$ level. This is perhaps most probable for the $4\tau$ final state since $c_{\tau\tau}$ is intuitively likely to be the largest of the $c$'s and $c_{\tau\tau}$ has no significant bounds at the moment.) In order to convert a measurement of or (more generally) limit on $BR(\Delta^{--} \to \ell^-\ell^-)$ for a given $\ell$ into a determination or bound on the corresponding $c_{\ell\ell}$, the total width, $\Gamma_{\Delta^{--}}$, of the $\Delta^{--}$ must be known. If $\Delta^{--} \to \Delta^-W^-$ is observed, the partial width $\Gamma(\Delta^{--} \to \Delta^-W^-)$ could be computed in any given model and combined with the $BR(\Delta^{--} \to \ell^-\ell^-)$ measurements and limits to determine $\Gamma_{\Delta^{--}}$. One would then be able to give model-dependent results/limits for the $c_{\ell\ell}$'s. If all the $c$'s are small, we would have only limits. The above procedure would fail if the $\Delta^{--} \to \Delta^-W^-$ final state has too small a branching ratio to be measured and $\Delta^{--}$ decays are dominated by one or more $\ell^-\ell^-$ final states. Without a direct measurement of $\Gamma_{\Delta^{--}}$, the magnitudes of the $c_{\ell\ell}$'s could not be determined (although, when more than one channel is seen, ratios could be obtained). The most that could be said is that the $c_{\ell\ell}$'s associated with the important decay channels would have to be large enough to overwhelm the $\Delta^{--} \to \Delta^-W^-$ decay channel. However, if this decay channel is not kinematically allowed, the semi-virtual $\Delta^{-*}W^-$ and $\Delta^-W^{-*}$ alternatives would have very tiny partial widths and this constraint would be satisfied for an enormous range of $c_{\ell\ell}$ values. Clearly, a direct and model-independent means for probing the $c_{\ell\ell}$ values regardless of size is needed.

# 3 Detecting the $\Delta^{--}$ in $e^-e^-$ Collisions

In this section,[2] we shall show that $e^-e^-$ collisions are capable of probing extremely small $c_{ee}$ values — values, for example, that span a very large portion of the parameter space for which the see-saw mechanism for neutrino mass would be natural.

A crucial ingredient in the potential of $e^-e^-$ collisions for producing the $\Delta^{--}$ at an observable rate is the $\sqrt{s}$ spectrum. This is determined by the amount of bremsstrahlung and beamstrahlung of photons from the initial $e^-$'s. Possible designs for the $e^-e^-$ collider are still being developed, but typically point to a spectrum that can be approximated by a Gaussian in the vicinity of the peak energy, with a 1 sigma rms resolution given by $\sigma \sim 0.2\% \sqrt{s}$, accompanied by a tail (coming from the beamstrahlung and bremsstrahlung). Current estimates [6] for a $250\,\text{GeV} \times 250\,\text{GeV}$ machine are that roughly 38% of the total luminosity will reside in the narrow Gaussian centered at the nominal machine energy, with the tail being such that the average energy loss from beamstrahlung and bremsstrahlung will be of order 3%. If the $e^-e^-$ collider is run at lower energies, more of the luminosity would remain in the central Gaussian peak. The instantaneous luminosity of the design now being considered is $\mathcal{L} \sim 6 \times 10^{33} cm^{-2} sec^{-1}$ for the $250\,\text{GeV} \times 250\,\text{GeV}$ case, leading to a total yearly luminosity of order $L = 60$ fb$^{-1}$, of which roughly $L = 25$ fb$^{-1}$ would reside in the central Gaussian peak. For the estimates made below, we adopt the working hypothesis that a few years of running will provide a total $L = 50$ fb$^{-1}$ in the central 0.2% Gaussian peak, for all machine energies below $\sqrt{s} \sim 500\,\text{GeV}$. Further, we shall ignore the extra luminosity that resides outside the Gaussian peak; this luminosity would act to increase the rate for $\Delta^{--}$ events, beyond the estimates to be given, when the $\Delta^{--}$ has a total width $\Gamma_{\Delta^{--}} \gtrsim 0.002 m_{\Delta^{--}}$.

A useful mnemonic for the Gaussian rms resolution, taking $\sqrt{s} = m_{\Delta^{--}}$, is

$$\sigma \sim 0.2 \,\text{GeV} \left(\frac{m_{\Delta^{--}}}{100\,\text{GeV}}\right) \left(\frac{R}{0.2\%}\right), \qquad (3.1)$$

where $R$ is the resolution in percent. The crucial issue is how $\sigma$ compares to $\Gamma_{\Delta^{--}}$. For $c_{\ell\ell} = 10^{-5}$ and $R = 0.2\%$, Eq. 2.3 predicts that $\Gamma(\Delta^{--} \to \ell^-\ell^-) = \sigma$ for $m_{\Delta^{--}} \sim 70\,\text{GeV}$. If all the $c'$ are much smaller than $10^{-5}$, the $\Delta^{--}$ is light, and the $\Delta^{--} \to \Delta^- W^-$ decay is either strongly suppressed or disallowed, then the $\Delta^{--}$ will have a width much smaller than $\sigma$. Conversely, if $m_{\Delta^{--}} \sim 500\,\text{GeV}$ and the $\Delta^- W^-$ decay channel has $\beta_{\Delta^- W^-} \geq 0.3$, then (even if all the $c_{\ell\ell}$'s are extremely small) $\Gamma_{\Delta^{--}} \geq 4.4\,\text{GeV}$, i.e. substantially larger than $\sigma \sim 1\,\text{GeV}$. We will present approximate results for $e^-e^- \to \Delta^{--}$ in the limits $\Gamma_{\Delta^{--}} \gg \sigma$ and $\Gamma_{\Delta^{--}} \ll \sigma$.

Using the Gaussian approximation, the effective cross section for $\Delta^{--}$ production in the $s$-channel is obtained by convoluting the standard $s$-channel pole form with the Gaussian distribution in $\sqrt{s}$ of rms width $\sigma$. The resulting cross section is denoted by $\overline{\sigma}_{\Delta^{--}}$. For $\Gamma_{\Delta^{--}} \gg \sigma$, $\Gamma_{\Delta^{--}} \ll \sigma$, $\overline{\sigma}_{\Delta^{--}}$ at $\sqrt{s} = m_{\Delta^{--}}$ is given by:

$$\overline{\sigma}_{\Delta^{--}} = \begin{cases} \frac{4\pi BR(\Delta^{--} \to e^-e^-)}{m^2_{\Delta^{--}}}, & \Gamma_{\Delta^{--}} \gg \sigma; \\ \frac{\sqrt{\pi}}{2\sqrt{2}} \frac{4\pi \frac{\Gamma(\Delta^{--} \to e^-e^-)}{\sigma}}{m^2_{\Delta^{--}}}, & \Gamma_{\Delta^{--}} \ll \sigma . \end{cases} \quad (3.2)$$

---

[2]Neutral Higgs detection via direct $s$-channel production in $\mu^+\mu^-$ collisions has been considered in Ref. [5]. This section employs some of the ideas developed there.

In terms of the integrated luminosity $L$, total event rates are given by $L\overline{\sigma}_{\Delta^{--}}$. As stated earlier, we will assume that $L = 50$ fb$^{-1}$ can be accumulated in the Gaussian peak centered at the nominal $e^-e^-$ energy.

Consider first the case where $\Gamma_{\Delta^{--}} \gg \sigma$. We find an event rate coming from the luminosity of the central Gaussian peak (the rate would actually be augmented in this case by a contribution from the beamstrahlung/bremsstrahlung tail) given by

$$N(\Delta^{--}) \sim 2.5 \times 10^{10} \left( \frac{100 \,\text{GeV}}{m_{\Delta^{--}}} \right)^2 BR(\Delta^{--} \to e^-e^-) . \quad (3.3)$$

If the $\Delta^{--} \to \Delta^- W^-$ decay mode dominates the total width, then $BR(\Delta^{--} \to e^-e^-) \sim 0.3 \beta^{-3}_{W^-\Delta^-}(c_{ee}/10^{-5})$. For $L = 50$ fb$^{-1}$ we would then have $3 \times 10^8$ $\Delta^{--}$ events (dominated by the $\Delta^- W^-$ final state) if $c_{ee} \sim 10^{-5}$, $\beta_{\Delta^- W^-} = 1$ and $m_{\Delta^{--}} = 500\,\text{GeV}$. As $\beta_{\Delta^- W^-}$ decreases below 1, the number of events grows rapidly. A total of 100 $\Delta^{--}$ events are produced for $c_{ee} = 1.3 \times 10^{-13}(m_{\Delta^{--}}/100\,\text{GeV})^2 \beta^3_{\Delta^- W^-}$, i.e. an observable signal would be present for incredibly small $c_{ee}$ values. We emphasize that the scenario of a large $\Delta^{--} \to \Delta^- W^-$ width is not so unlikely. If the $\tau\tau$ decay mode dominates $\Gamma_{\Delta^{--}}$, then $BR(\Delta^{--} \to e^-e^-) \sim c_{ee}/c_{\tau\tau}$. For $L = 50$ fb$^{-1}$, $10^8$ $\Delta^{--}$ events (almost entirely $\tau^-\tau^-$) would be obtained for $c_{ee}/c_{\tau\tau} = 0.1$ and $m_{\Delta^{--}} = 500\,\text{GeV}$. In this case, 100 $\Delta^{--}$ events would correspond to $c_{ee}/c_{\tau\tau} = 4 \times 10^{-9}(m_{\Delta^{--}}/100\,\text{GeV})^2$, again a very respectable sensitivity. Note that the phenomenology of this latter case of $\tau^-\tau^-$ dominance of $\Delta^{--}$ decays is essentially independent of $\Gamma(\Delta^{--} \to \Delta^- W^-)$.

In this $\Gamma_{\Delta^{--}} \gg \sigma$ case, it is important to note that a measurement or calculation of $\Gamma_{\Delta^{--}}$ is required in order that the value of $BR(\Delta^{--} \to e^-e^-)$ determined from $N(\Delta^{--})$ (see Eq. 3.3) and/or direct observation can be used to compute $\Gamma(\Delta^{--} \to e^-e^-)$ and, thence, $c_{ee}$. A calculation of $\Gamma(\Delta^{--} \to \Delta^- W^-)$ is possible given a specific choice of the Higgs representation and the observational knowledge of the masses $m_{\Delta^{--}}$ and $m_{\Delta^-}$. If $\Delta^{--} \to \Delta^- W^-$ is the dominant decay mode, then this yields a fairly accurate value for $\Gamma_{\Delta^{--}}$. However, if the Higgs representation is not known, or $\Delta^{--} \to e^-e^-$, $\mu^-\mu^-$, and/or $\tau^-\tau^-$ decays dominate to such an extent that $BR(\Delta^{--} \to \Delta^- W^-)$ cannot be extracted from the data, then determination of $\Gamma_{\Delta^{--}}$ will require its direct measurement. We return to this issue shortly.

The other useful benchmark scenario is that in which $\Delta^{--} \to \Delta^- W^-$ is either highly suppressed or forbidden, and all of the $c_{\ell\ell}$'s are relatively small. In this case, $\Gamma_{\Delta^{--}} \ll \sigma$ is probable, with *very* narrow widths being predicted if the $\Delta^- W^-$ mode is forbidden. Taking $L = 50$ fb$^{-1}$, and using Eq. 3.1 for $\sigma$ and the result in Eq. 2.3 for $\Gamma(\Delta^{--} \to e^-e^-)$, we find from Eq. 3.2 an event rate of

$$N(\Delta^{--}) \sim 3 \times 10^{10} \left(\frac{c_{ee}}{10^{-5}}\right) \left(\frac{0.2\%}{R}\right) ; \qquad (3.4)$$

clearly an enormous event rate results if $c_{ee}$ is within a few orders of magnitude of its upper bound. If the $\Delta^- W^-$ decay is two-body allowed but all $c_{\ell\ell}$'s are very small, the $\Delta^{--}$ final state would be dominated by the real $\Delta^- W^-$ mode (even though $\beta_{\Delta^- W^-} \ll 1$); if $\Delta^- W^-$ is two-body forbidden, one or several of the $\ell^-\ell^-$ modes would dominate unless all the $c_{\ell\ell}$'s are extremely small, in which case the $\Delta^{-*}W^-$, $\Delta^- W^{-*}$ semi-virtual three-body modes would be dominant. The precise cross-over point between the $\ell^-\ell^-$ modes and the semi-virtual modes depends on details and will not be pursued here. (Some sample scenarios illustrating this cross-over were explored in Ref. [3].) Note that if the $\Delta^{--}$ is observed at the LHC or NLC, we will know ahead of time what final state to look in and its detailed characteristics, even if the semi-virtual final state is dominant. Only the latter semi-virtual modes and the $e^-e^-$ final state would have significant backgrounds at an $e^-e^-$ collider.

We emphasize that, in the $\Gamma_{\Delta^{--}} \ll \sigma$ case, Eq. 3.4 shows that the event rate alone is sufficient to determine $c_{ee}$, unlike in the $\Gamma_{\Delta^{--}} \gg \sigma$ case. Direct measurement of $\Gamma_{\Delta^{--}}$ is not required, but would yield important additional information, as described shortly.

To estimate our ultimate sensitivity to $c_{ee}$ when $\Gamma_{\Delta^{--}} \ll \sigma$, let us suppose that 100 events are required for observation in the real $\Delta^- W^-$ and $\ell^-\ell^-$ final state scenarios, and 1000 events if the semi-virtual final states dominate. From Eq. 3.4, we predict 100 $\Delta^{--}$ events for $c_{ee} \sim 3.3 \times 10^{-14}(R/0.2\%)$; note that this result does not depend upon $m_{\Delta^{--}}$. Once again, we have dramatic sensitivity. Even in the worst case scenario of requiring 1000 events when the semi-virtual modes dominate the final state, we are able to achieve a nearly 8 orders of magnitude improvement over the current limits on $c_{ee}$. Due to the large direct $e^-e^- \to e^-e^-$ background, $\gtrsim 1000$ events might also be required if $e^-e^-$ final states dominated the $\Delta^{--}$ decay. However, it seems rather likely that $c_{ee} < c_{\mu\mu}$ and $c_{\tau\tau}$, in which case this situation would not arise. If the $\mu^-\mu^-$ final state were dominant, as few as 10 events would probably constitute a viable signal.

In practice, the LHC determination of $m_{\Delta^{--}}$ in $\Delta^{--}\Delta^{++}$ pair production will be imperfect. This is not too important if $\Gamma_{\Delta^{--}}$ is large, but could be a significant factor if $\Gamma_{\Delta^{--}} < \sigma$ since then a limited scan would become necessary in order to be certain that at least one energy setting corresponded to $\sqrt{s} \simeq m_{\Delta^{--}}$ to within a fraction of $\sigma$. In $\Delta^{--}\Delta^{++}$ production, the smallest error, $\delta m_{\Delta^{--}}$ for $m_{\Delta^{--}}$ will be achievable in $4e$ or $4\mu$ final states. The worst case scenario would be dominance of $\Delta^{--} \to \Delta^- W^-$ decays coupled with a very narrow partial width (due to $\beta_{\Delta^- W^-} \ll 1$). The minimum $c_{ee}$ for which we will be able to detect the $\Delta^{--}$ increases proportionally to the number of scan points required to span $2\delta m_{\Delta^{--}}$ at intervals of $\sim \sigma$. One could conceivably lose as much as a factor of 10 in $c_{ee}$ sensitivity in some cases.

In the very unlikely event that we are unable to exclude the existence of a $\Delta^{--}$ with $m_{\Delta^{--}} \lesssim 500\,\text{GeV}$ by searching for $\Delta^{--}\Delta^{++}$ pairs at the LHC, then directly searching for a $\Delta^{--}$ at the $e^-e^-$ collider could be considered. This would require scanning over a broad energy range. Assuming that we could be confident from the NLC that there is no $\Delta^{--}$ with mass below about $250\,\text{GeV}$, then for $R = 0.2\%$ we would need about 350-400 energy settings to cover the $250 - 500\,\text{GeV}$ mass range. This would obviously increase the minimum $c_{ee}$ value for which a signal could be detected at each scan point by a similar factor. However, if the $\Delta^{--}$ has a large $\Gamma_{\Delta^{--}}$, a signal would emerge for smaller $c_{ee}$ by combining individual scan points. If the $\Delta^{--}$ is very narrow, the smallest possible $R$ value coupled with a finer scan would maximize the chance of seeing a signal in the $\Delta^{-*}W^-$ and $e^-e^-$ channels for which the background is significant.

Smaller $R$ would also allow a direct measurement, by scanning, of smaller $\Gamma_{\Delta^{--}}$. Measurement of $\Gamma_{\Delta^{--}}$ would provide very important additional information regarding the $\Delta^{--}$ in many of the partial width scenarios we have described, but, of course, might also be simply impossible if $\Gamma_{\Delta^{--}}$ is very tiny. As noted earlier, the case in which a direct $\Gamma_{\Delta^{--}}$ determination by scanning would be most important is that in which $\Gamma_{\Delta^{--}} > \sigma$, since the magnitude of $\Gamma_{\Delta^{--}}$ is needed in order to convert the value of $BR(\Delta^{--} \to e^-e^-)$ into a determination of $c_{ee}$. Fortunately, when $\Gamma_{\Delta^{--}} > \sigma$ measurement of the $\Delta^{--}$ total width by scanning will be straightforward. In contrast, if $\Gamma_{\Delta^{--}} \ll \sigma$ we find from Eq. 3.4 that $N(\Delta^{--})$ provides a direct determination of $c_{ee}$. This is a crucial fact given that a very small $\Gamma_{\Delta^{--}}$ might not be measurable directly. In general, if $\Gamma_{\Delta^{--}}$ is known, values for $BR(\Delta^{--} \to \ell^-\ell^-)$ for any of the $\ell = e, \mu, \tau$ channels will then allow us to determine the corresponding $c_{\ell\ell}$. This applies, in particular, for $\ell = \mu, \tau$ in the $\Gamma_{\Delta^{--}} \ll \sigma$ case where event rate alone is adequate to determine $c_{ee}$. It is important to note that direct measurement of $\Gamma_{\Delta^{--}}$ combined with measurement of the $\mu^-\mu^-$ and $\tau^-\tau^-$ branching ratios is the only means for determining $c_{\mu\mu}$ and $c_{\tau\tau}$ in $e^-e^-$ collisions. We will not discuss here the luminosity required for determining $\Gamma_{\Delta^{--}}$ by scanning except to note that it will increase rapidly as $\Gamma_{\Delta^{--}}$ decreases below $\sigma$. The simple process of centering $\sqrt{s}$ to a value $\simeq m_{\Delta^{--}}$ will already have provided a first crude measurement of $\Gamma_{\Delta^{--}}$ if $\Gamma_{\Delta^{--}}$ is not too much smaller than $\sigma$.

Before concluding this section, we note that the ability to polarize the beams could prove very valuable. Returning to Eq. 1.1, we see that the hypercharge of the $\Delta^{--}$ could be determined directly (up to possible extensions of the charge formula such as in the left-right symmetric models), not to mention the fact that the cross section would be enhanced by a factor of four.

# 4 Final Remarks and Conclusions

Although currently out of favor because of the success of the minimal supersymmetric model, there are well-motivated models containing triplet and other Higgs representations which include a $\Delta^{--}$ Higgs boson. If such a boson exists in the mass range $\lesssim 500\,\text{GeV}$ accessible to the $e^-e^-$ collider option at the NLC, it is very likely to be observed at the LHC, even if too heavy ($\gtrsim 240\,\text{GeV}$) to be seen in pair production in $e^+e^-$ collisions at the NLC. If a $\Delta^{--}$ is detected at either the NLC or LHC, we have demonstrated that the $e^-e^-$ collider could be employed as a $\Delta^{--}$ factory, producing potentially billions of $\Delta^{--}$'s per year if the $e^-e^- \to \Delta^{--}$ coupling is near its current upper bound. More generally, limits on this lepton-number-violating coupling could be improved by roughly 8 orders of magnitude at the $e^-e^-$ collider, with some dependence on the $\Delta^{--}$ total width and decay pattern. In left-right symmetric models, this implies sensitivity to much of the coupling strength range for which the see-saw mechanism for neutrino mass generation operates most naturally.

Further, if the total width, $\Gamma_{\Delta^{--}}$, of the $\Delta^{--}$ can be measured by scanning, and if a given $\ell^-\ell^-$ final state has measurable branching ratio, then we can combine these quantities to obtain the $\ell^-\ell^- \to \Delta^{--}$ coupling. This is the only technique for determining this coupling in the $\ell = \mu, \tau$ cases. It is also the only way to directly determine the $e^-e^-$ coupling when $\Gamma_{\Delta^{--}}$ is larger than the $\sqrt{s}$ resolution.

We emphasize that if $m_{\Delta^{--}} \gtrsim \sqrt{s}/2$ for the $e^+e^-$ collider, then in order to avoid a broad scan search for a $\Delta^{--}$ we must have an approximate determination of $m_{\Delta^{--}}$ via detection of $\Delta^{--}\Delta^{++}$ pair production at the LHC. (Such a determination would be especially crucial if $\Gamma_{\Delta^{--}}$ is much smaller than the energy resolution $\sigma$ of the $e^-e^-$ collider.) This fact provides yet another example of the complementarity of the NLC and the LHC.

Of course, an exactly parallel set of results would apply to a $\mu^-\mu^-$ collider.[3] Indeed, there are two potential advantages of a $\mu^-\mu^-$ collider over an $e^-e^-$ collider. Both advantages derive from the much reduced beamstrahlung and bremsstrahlung associated with muon beams. First, the resolution $R$ for a $\mu^-\mu^-$ collider could potentially be much smaller than that for an $e^-e^-$ collider, and, in addition, more of the total luminosity will reside in the Gaussian peak centered at the nominal machine energy. Preliminary studies of $\mu^+\mu^-$ colliders indicate that $R$ values as small as $R \sim 0.01\%$ might be achievable [8]. In the case that $\Gamma_{\Delta^{--}}$ is very small, a factor of roughly $R(\mu\mu)/R(ee)$ increase in the $\Delta^{--}$ production rate and corresponding sensitivity to $c_{\mu\mu}$ vs. $c_{ee}$ would result from the superior resolution alone. In the absence of on-shell $\Delta^-W^-$ decays of the $\Delta^{--}$, $c_{\mu\mu}$ values in the $10^{-15}$ range would be probed for $R(\mu\mu) \sim 0.01\%$ assuming that the $\Delta^{--}$ is already discovered at the LHC or NLC so that a broad scan is not necessary. The second advantage of a $\mu\mu$ collider might turn out to be larger energy reach. It is anticipated [7] that $\sqrt{s}$ as high as 4 TeV might eventually prove feasible. Thus, $\Delta^{--}\Delta^{++}$ pair production could be detected in the $\mu^+\mu^-$ mode of operation up to very high $m_{\Delta^{--}}$, and then the $\mu^-\mu^-$ mode of operation would allow a high-sensitivity probe of $c_{\mu\mu}$. Of course, if the $\Delta^{--}$ could be detected in both $e^-e^-$ and $\mu^-\mu^-$ collisions, then we would measure both $c_{ee}$ and $c_{\mu\mu}$. These determinations of $c_{ee}$ and $c_{\mu\mu}$ could then be compared to those independently extracted by measuring $\Gamma_{\Delta^{--}}$, $BR(\Delta^{--} \to e^-e^-)$ and $BR(\Delta^{--} \to \mu^-\mu^-)$ at the $e^-e^-$ and/or $\mu^-\mu^-$ colliders.

---

[3]We note that $\mu^+\mu^-$ colliders are already being actively considered [7].

Overall, it is apparent that if a doubly-charged Higgs boson is found at the NLC or LHC, $e^-e^-$ and $\mu^-\mu^-$ colliders would separately and in combination provide enormously important information concerning the structure and interactions of the Higgs sector.

## 5 Acknowledgements

This work was supported in part by Department of Energy grant DE-FG03-91ER40674 and the Davis Institute for High Energy Physics.